%% file: main.tex
\documentclass[
]{ceurart}

\usepackage{listings}
\usepackage{tikz}
\usepackage{subcaption}
\begin{document}

\copyrightyear{2025}
\copyrightclause{Copyright for this paper by its authors.
  Use permitted under Creative Commons License Attribution 4.0
  International (CC BY 4.0).}

\conference{Proceedings of AI4HGI ’25, the First Workshop on Artificial Intelligence for Human-Game Interaction at the 28th European Conference on Artificial Intelligence (ECAI ’25), Bologna, October 25-30, 2025}

\title{Diamonds in the rough: Transforming SPARCs of imagination into a game concept by leveraging medium sized LLMs}


\author[1]{Julian Geheeb}[%
orcid=0009-0006-9607-7548,
email=julian.geheeb@tum.de,
]
\cormark[1]
\address[1]{Technical University of Munich, Arcisstraße 21, 80333 Munich, Germany}

\author[1]{Farhan Abid Ivan}[
email=farhanabid.ivan@tum.de,
]

\author[1]{Daniel Dyrda}[
orcid=0000-0002-0394-3325,
email=daniel.dyrda@tum.de,
]

\author[1]{Miriam Anschütz}[
orcid=0009-0009-8487-9481,
email=miriam.anschuetz@tum.de,
]

\author[1]{Georg Groh}[
orcid=0000-0002-5942-2297,
email=grohg@cit.tum.de,
]

\cortext[1]{Corresponding author.}

\begin{abstract}
  Recent research has demonstrated that large language models (LLMs) can support experts across various domains, including game design. In this study, we examine the utility of medium-sized LLMs—models that operate on consumer-grade hardware typically available in small studios or home environments.
  We began by identifying ten key aspects that contribute to a strong game concept and used ChatGPT to generate thirty sample game ideas. Three medium-sized LLMs—LLaMA 3.1, Qwen 2.5, and DeepSeek-R1—were then prompted to evaluate these ideas according to the previously identified aspects. A qualitative assessment by two researchers compared the models' outputs, revealing that DeepSeek-R1 produced the most consistently useful feedback, despite some variability in quality.
  To explore real-world applicability, we ran a pilot study with ten students enrolled in a storytelling course for game development. At the early stages of their own projects, students used our prompt and DeepSeek-R1 to refine their game concepts. The results indicate a positive reception: most participants rated the output as high quality and expressed interest in using such tools in their workflows.
  These findings suggest that current medium-sized LLMs can provide valuable feedback in early game design, though further refinement of prompting methods could improve consistency and overall effectiveness.
\end{abstract}


\maketitle

\input{files/introduction.tex}

\input{files/framework.tex}

\input{files/experimental_setup.tex}

\input{files/model_comparison.tex}

\input{files/sparc.tex}

\input{files/discussion.tex}

\input{files/related_work.tex}

\input{files/conclusion.tex}

\begin{acknowledgments}
  Thanks to the developers of ACM consolidated LaTeX styles
  \url{https://github.com/borisveytsman/acmart} and to the developers
  of Elsevier updated \LaTeX{} templates
  \url{https://www.ctan.org/tex-archive/macros/latex/contrib/els-cas-templates}.  
\end{acknowledgments}

\section*{Declaration on Generative AI}
During the preparation of this work, the authors wrote a full draft of the paper and subsequently used chatgpt.com with GPT-4o to improve the writing style and grammar. Further, the authors used perplexity.ai to get an initial overview of related papers. After using these tools/services, the authors reviewed and edited the content as needed and take full responsibility for the publication’s content. 

\bibliography{bibliography}

\end{document}

%% file: files/introduction.tex
\section{Introduction}\label{sec:introduction}
At the beginning of any creative process, there is often a spark—a moment of imagination that captures an idea and sets it on a path toward becoming a finished artifact. In our context, this artifact is a video game. However, the initial concept is typically rough and underdeveloped, like an unpolished gem. It requires refinement before it can serve as a foundation for development.

This refinement begins in the pre-production stage of game development, particularly during the conceptualization phase, where the core idea is expanded into a full-fledged game concept~\cite{kanode2009software}. To be effective, such a concept must include a sufficient level of detail across various dimensions, enabling smoother transitions into later stages of production. Because these concepts are often documented in written formats—such as Game Design Documents (GDDs)—large language models (LLMs) present a promising tool for evaluating whether this level of detail has been achieved.

LLMs have demonstrated their ability to support experts across many fields~\cite{nazi2024large, luo2025large}, including game design~\cite{lanzi2023chatgpt}. However, most high-performance LLMs require significant computational resources and are typically accessed via cloud-based platforms. This reliance on third-party providers introduces concerns about privacy, intellectual property, and long-term accessibility—issues particularly relevant to independent game developers and small studios.

In this study, we explore whether medium-sized LLMs, which can be hosted on consumer-grade hardware, can provide meaningful support during the conceptualization phase of game design. Specifically, we investigate whether these models can deliver valuable feedback on early-stage game concepts without the need for external cloud services.

To address this question, our contributions are as follows:
\begin{itemize}
\item We identify ten key aspects that characterize a robust game concept (\autoref{sec:framework}).
\item We conduct a human evaluation to compare three medium-sized models using a test dataset, our ten key aspects, and standardized hardware (\autoref{sec:model_comparison}).
\item We build a prototype, SPARC, and run a pilot study in which the best-performing model is integrated into the workflow of students engaged in early-stage game development (\autoref{sec:sparc}).
\end{itemize}

%% file: files/framework.tex
\section{Conceptualization Framework}\label{sec:framework}
To enable the models to meaningfully evaluate game concepts, we first defined a set of criteria grounded in established game development practices. Drawing from a range of sources in game design, level design, and production—such as Salen and Zimmerman~\cite{tekinbas2003rules}, Schell~\cite{schell2008art}, Galuzin~\cite{galuzin2016preproduction}, Totten~\cite{totten2017level}, Fullerton~\cite{fullerton2024game}, and Yang~\cite{leveldesignbook}—we identified ten key aspects that a well-developed game concept should address.
While not exhaustive, these aspects offer a solid foundation for evaluating early-stage design ideas and are well suited to the aims of our study. In practical settings, they could also be adapted to the specific needs of individual teams or projects.
Each aspect was carefully defined in an extended description, which was included in the prompt provided to the LLMs. A brief overview of the ten aspects is presented below.

\paragraph{Player Experience}
This aspect describes what the player is supposed to experience. It is written from the perspective of the player in the active form focusing on emotional experiences and it should include a high concept statement for the play idea.

\paragraph{Theme}
This aspect defines the theme of the idea. The theme of a game concept is often divided into a dominant unifying theme and multiple secondary themes.

\paragraph{Gameplay}
This aspect describes the core gameplay. It includes finding 3--5 verbs that describe the gameplay experience and it should include a \textit{30 seconds of gameplay} statement describing what the player typically does.

\paragraph{Place}
This aspect defines places in the game world where the space under construction can be set. It includes the environment setting of the idea, which is similar to theme, but it describes an actual location within the game world. This aspect should also provide a list of concrete locations the game takes place in.

\paragraph{Unique Features}
This aspect consists of a list with 3-5 features that are the defining elements of the idea. It answers the question of how the idea will be unique by contrasting it to existing projects.

\paragraph{Story and Narrative}
This aspect describes the rough story of the game and how the player experiences it. It includes defining storytelling methods, such as environmental storytelling, gameplay, cutscenes, narrators, dialogues, story context, and more.

\paragraph{Goals, Challenge and Rewards}
This aspect defines goals, challenges and rewards for the idea. Goals define objectives that the player has to complete. Challenges are obstacles the player has to overcome in order to achieve one goal. The rewards describe how the player will be rewarded for overcoming a set of obstacles to achieve one goal.

\paragraph{Art Direction}
This aspect describes the general artistic vision. It should include an art style, color palettes, and visually unique features.

\paragraph{Purpose}
This aspect defines the purpose of the project. It includes formulating the purpose for all involved stakeholders on why they want to work on the project.

\paragraph{Opportunities and Risks}
This aspect describes opportunities and risks of the idea by providing a list of each. For the opportunities, it includes planning on how to use them effectively. For the risks, it includes how likely they are to happen and strategies to minimize the risks.

%% file: files/experimental_setup.tex
\section{Hardware Setup}\label{sec:hardware}
As outlined in \autoref{sec:introduction}, our objective was to ensure that the proposed approach remains accessible to small indie developers and hobbyists by relying on locally available consumer-grade hardware. 
To this end, we selected a representative system configuration that served as the baseline for our study (see the left side of \autoref{tab:system_configuration}).
All model selections and experiment designs were made with this system’s capabilities in mind, ensuring that the approach is technically feasible on such hardware. 
We chose to execute the non--user-facing experiments in \autoref{sec:model_comparison} on a more powerful machine to reduce runtime (see the right side of \autoref{tab:system_configuration}), but the system detailed on the left represents the minimum specifications required to reproduce the methodology.

\vspace{5mm}

\begin{table}[ht]
    \centering
    \begin{tabular}{ll}
        \toprule
        \multicolumn{2}{c}{\textbf{Baseline System Configuration}} \\
        \midrule
         Operating System & Ubuntu 22.04 \\
         GPU & NVIDIA GeForce RTX 3080 Ti \\
         Memory & 12 GB of GDDR6X \\
        \bottomrule
    \end{tabular}
    \hspace{1cm}
    \begin{tabular}{ll}
        \toprule
        \multicolumn{2}{c}{\textbf{Faster System Configuration}} \\
        \midrule
         Operating System & Ubuntu 22.04 \\
         GPU & 2× NVIDIA A40 \\
         Memory & 48 GB of GDDR6 \\
        \bottomrule
    \end{tabular}
    \caption{Baseline system configuration (left), used to guide model and experiment design, and faster system configuration (right), used to reduce runtime for experiments in \autoref{sec:model_comparison}.}
    \label{tab:system_configuration}
\end{table}

%% file: files/model_comparison.tex
\section{Model Comparison and Qualitative Analysis}\label{sec:model_comparison}
This section outlines the methodology, results, and discussion of our first experiment, in which we compared the outputs of three medium-sized LLMs: \textit{meta-llama/Llama-3.1-8B-Instruct\footnote{\href{https://huggingface.co/meta-llama/Llama-3.1-8B-Instruct}{https://huggingface.co/meta-llama/Llama-3.1-8B-Instruct}}}~\cite{meta2025llama31}, \textit{Qwen/Qwen2.5-7B-Instruct}\footnote{\href{https://huggingface.co/Qwen/Qwen2.5-7B-Instruct}{https://huggingface.co/Qwen/Qwen2.5-7B-Instruct}}~\cite{team2024qwen2}, and \textit{deepseek-ai/DeepSeek-R1-Distill-Llama-8B\footnote{\href{https://huggingface.co/deepseek-ai/DeepSeek-R1-Distill-Llama-8B}{https://huggingface.co/deepseek-ai/DeepSeek-R1-Distill-Llama-8B}}}~\cite{guo2025deepseek} (\textit{LLaMA 3.1}, \textit{Qwen 2.5}, and \textit{DeepSeek-R1} in the following).
All three models were selected based on their compatibility with the baseline system described in \autoref{sec:hardware}, though the actual evaluation was conducted on a more powerful machine to expedite processing. This section focuses on the comparative analysis itself; hardware-specific execution details are discussed in their relevant context.

\subsection{Methodology}
The comparison was conducted through a qualitative human evaluation involving two researchers. To enable this, we first created a custom dataset of game ideas and collected model outputs for each entry. The following subsections describe both the data generation process and the evaluation procedure in detail.

\input{files/game_data_set.tex}

\input{files/model_prompting.tex}

\input{files/human_evaluation/method.tex}

\input{files/human_evaluation/results.tex}

\subsection{Discussion}
This section discusses the model comparison and qualitative evaluation, with a focus on informing the implementation of the prototype system described in \autoref{sec:sparc}. Broader implications and general reflections are addressed separately in \autoref{sec:discussion}.

Our aim was to compare the performance of LLaMA 3.1, Qwen 2.5, and DeepSeek-R1 in providing structured feedback on game concepts. In our experimental setup, both LLaMA 3.1 and Qwen 2.5 failed to consistently adhere to the required output format and completeness criteria. While alternative prompting strategies or setups might potentially improve their performance, we chose to focus our in-depth analysis on DeepSeek-R1, which showed the most promise in terms of structural consistency and coverage.

DeepSeek-R1 reliably produced outputs structured around the ten predefined aspects of a game concept—an important requirement for our goal of enabling systematic feedback. Although the quality of feedback varied across individual outputs, the model’s ability to maintain structural coherence and generally relevant content led us to proceed with prototype development and a subsequent pilot study. In short, the model demonstrated sufficient capability to warrant practical exploration.

An additional consideration emerged regarding the dataset used during model evaluation. While dataset design was not the primary focus of this phase, we observed a recurring bias in ChatGPT toward large-scale or high-concept game ideas, which may not reflect the scope or constraints of smaller indie studios. Many ideas shared recurring tropes—such as the presence of multiple coexisting dimensions in space or time—which may reflect limitations of the generation prompt or the model’s training data.

These biases, while not critical for the current phase, should be addressed in future iterations—particularly in the context of the pilot study, where feedback will be applied to participants’ own early-stage game concepts. This shift from synthetic to authentic data will allow for more targeted evaluation of model utility in real-world design contexts.

%% file: files/game_data_set.tex
\subsubsection{Game Idea Dataset Creation}\label{sec:idea_generation}
To evaluate the capabilities of different language models and enable consistent comparisons, we first created a dataset of game ideas with varying levels of descriptive detail. We used OpenAI’s GPT-4o~\cite{hurst2024gpt}, accessed through its official chat interface\footnote{\href{https://chatgpt.com/}{https://chatgpt.com/}}, to generate both the ideas and corresponding summaries.
\begin{figure}
    \centering
    \includegraphics[trim=0 0 75 0, clip, width=\textwidth]{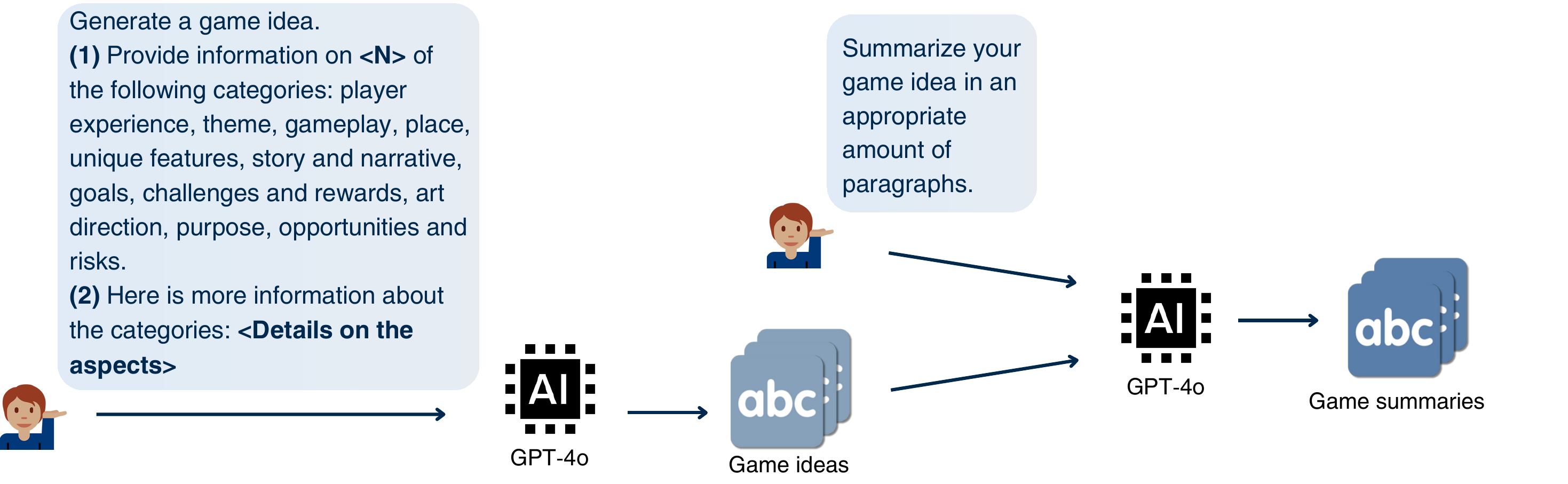}
    \caption{Depiction of the prompts and workflow used to generate the test dataset. The left prompt has additional options (1) and (2) used to refine the process.}
    \label{fig:game-idea-generation}
\end{figure}
The generation process followed these steps:
\begin{itemize}
    \item Prompt GPT-4o to generate a game idea (left prompt in \autoref{fig:game-idea-generation}), optionally specifying how many aspects to cover (1) and whether to include detailed descriptions of those aspects (2).
    \item Save the generated game idea as a plain text file for further processing.
    \item Prompt GPT-4o to produce a summary of the same idea (right prompt in \autoref{fig:game-idea-generation}).
    \item Save the summary in a separate text file for evaluation.
\end{itemize}

\paragraph{Results}
Using the first prompt shown in \autoref{fig:game-idea-generation}, we generated 15 distinct game ideas under varying conditions to ensure diversity in content and coverage. Each idea included both a full version and a summary, resulting in a total of 30 text files.
This structured variety allowed us to test model performance across a range of input detail levels while maintaining consistency in generation logic.
The prompt configurations were as follows:

\begin{itemize}
\item 4 game ideas generated without options (1) or (2),
\item 5 game ideas using option (1) only, with randomly selected values for <N>: $[3, 5, 7, 7, 8]$,
\item 3 game ideas using both options (1) and (2), with randomly selected values for <N>: $[6, 6, 8]$,
\item 3 game ideas using both options (1) and (2), with <N> fixed at 10.
\end{itemize}

Therefore, an example prompt of the second configuration would be as follows, where the selection of categories was determined by the LLM: 
\begin{quote}
Generate a game idea. Provide information on three of the following categories: player experience, theme, gameplay, place, unique features, story and narrative, goals, challenges and rewards, art direction, purpose, opportunities and risks.
\end{quote}

%% file: files/model_prompting.tex
\subsubsection{Model Prompting}
For the model comparison, we used a standardized evaluation prompt (\autoref{fig:final_prompt}) across all models tested. This prompt instructed each model to assess whether the key aspects required for initiating game development were present or inferable in each game idea.
To generate outputs for all 30 game ideas from \autoref{sec:idea_generation}, we employed the Hugging Face Text Generation Inference Docker\footnote{\href{https://huggingface.co/docs/text-generation-inference/en/index}{https://huggingface.co/docs/text-generation-inference/en/index}}. This environment streamlined inference across various open-source LLMs, including the three models selected for our comparison.
Each model was prompted once per game idea, resulting in a total of 90 output files. Although the models were chosen for their compatibility with the baseline system, this phase of the experiment was executed on a more powerful system, as discussed in \autoref{sec:hardware} and shown in \autoref{tab:system_configuration}.

\begin{figure}
    \centering
    \begin{tikzpicture}
        \node[
            draw,
            fill=orange!10,
            rounded corners,
            text width=13cm,
            inner sep=10pt,
            align=justify
        ] {You are an expert game development consultant. Your task is to evaluate the following game text as the foundation for a game development project. 
        Check if the following aspects are present or can be easily inferred from the game idea: player experience, theme, gameplay, place, unique features, story and narrative, goals, challenges and rewards, art direction, purpose, opportunities and risks. 
        Expanded details about the aspects are as follows:
        \newline
        \textbf{<Details on the aspects>}
        \newline
        The objective is to check whether fields and aspects required to start development of a game have been considered. 
        Add suggestions at the end of evaluation along with 2-5 other details that would make the text better suited to start game development with in addition to including aspects that are not addressed in the game text. 
        Do not take into account fiscal or managerial requirements. Focus only on factors relevant for early stages of game design. 
        Avoid redundancy and limit your response to 1000 words.};
    \end{tikzpicture}
    \caption{Prompt used to generate structured evaluation outputs for all models. The placeholder <Details on the aspects> corresponds to content adapted from \autoref{sec:framework}, which has been omitted here for brevity. Full details are available upon request.}
    \label{fig:final_prompt}
\end{figure}

%% file: files/human_evaluation/method.tex
\subsubsection{Human Evaluation}\label{sec:human_evaluation_method}
Following the generation of model outputs, we conducted a two-phase human evaluation to assess the structure, relevance, and quality of the responses.
The evaluation was conducted independently by two researchers who were also closely involved in defining the ten aspects outlined in \autoref{sec:framework}.

The first phase involved a high-level comparison of the 90 outputs (30 game ideas × 3 models) to determine whether each model was capable of providing structured and usable feedback. This phase aimed to answer the overarching question:
\textit{Can this model provide structured and coherent feedback on game concepts?}
The outputs were evaluated against the following general criteria:
\begin{itemize}
    \item \textbf{Format}: Does the response follow the requested structure and formatting?
    \item \textbf{Completeness}: Does the model address all ten predefined aspects?
    \item \textbf{Clarity and Coherence}: Is the language clear, and does the feedback make logical sense overall?
\end{itemize}

The second phase focused on a closer qualitative review of the 30 outputs generated by the model selected as most promising in the first phase. This detailed assessment combined open-ended analysis with the following targeted criteria:
\begin{itemize}
    \item \textbf{Comprehension}: Does the model correctly interpret the game idea and identify the relevant aspects?
    \item \textbf{Specificity}: Is the feedback tailored to the individual game idea, or is it overly generic?
    \item \textbf{Hallucination}: To what extent does the model introduce unfounded or invented content?
    \item \textbf{Feedback Quality}: How valuable and well reasoned is the feedback from a game design perspective?
\end{itemize}

This two-phase process allowed us to first filter for viability and then examine depth and reliability in greater detail.

%% file: files/human_evaluation/results.tex
\subsection{Results - Phase 1: Comparative Evaluation Across Models}
\begin{table}[ht]
    \centering
    \begin{tabular}{lccc}
        \toprule
        & \textbf{Format} & \textbf{Completeness} & \textbf{Clarity}\\
        \midrule
        \textbf{LLaMA 3.1} & 1/30 & 20/30 & 30/30 \\
        \textbf{Qwen 2.5} & 1/30 & 30/30 & 30/30 \\
        \textbf{DeepSeek-R1} & 30/30 & 26/30 & 27/30 \\
        \bottomrule
    \end{tabular}
    \caption{Evaluation results for phase~1. The table reports, for each model, the number of outputs (out of 30) that satisfied the criteria of format, completeness, and clarity.}
    \label{tab:phase_1_results}
\end{table}

In terms of format, DeepSeek-R1 consistently outperformed the other two models (see \autoref{tab:phase_1_results}).
Outputs from LLaMA 3.1 and Qwen 2.5 frequently entered infinite loops, repeating the last sentence, paragraph, or entire structure until reaching the maximum token limit.
These outputs were typically cut off mid-sentence once the limit was reached, and they often failed to follow the structured format specified in the prompt—namely, organizing feedback around the ten predefined aspects.

In contrast, DeepSeek-R1 never exhibited looping behavior and provided structured feedback covering all ten aspects in 26 out of 30 cases. 
However, a minor issue was observed in 3 out of 30 outputs, where the model produced unexpected language artifacts, inserting Chinese characters mid-sentence. 
Despite this, clarity and coherence were generally comparable across all models—aside from the looping and formatting issues, no major qualitative differences were noted in this category.

\subsection{Results — Phase 2: In-Depth Analysis of DeepSeek Outputs}
Given its strong performance in Phase~1, DeepSeek-R1 was selected for a more detailed analysis in Phase~2. 
We observed two distinct output structures across the model's responses:
\begin{itemize}
    \item \textbf{Summary-first structure} — the model begins by summarizing the original game idea according to the ten aspects, followed by a set of suggestions and feedback.
    \item \textbf{Integrated structure} — feedback and suggestions are embedded directly within each aspect's analysis, creating a more intertwined and iterative review.
\end{itemize}
The integrated structure typically focused on feedback, while the summary-first structure emphasized summarization. This distinction made the two structures clearly recognizable in our observations.
In practice, many outputs exhibited variations or hybrid forms of these two patterns, but nearly all could be classified within or between these structural types, which were approximately evenly distributed.

The depth and detail of feedback varied significantly across different game ideas. In general, the model tended to echo the aspects explicitly stated in the prompt rather than invent new ones—indicating a low level of hallucination. The model found this easier to do with the original game idea compared to its summary. However, there were occasional instances where speculative ideas were presented as factual. For example, the model sometimes introduced key locations not mentioned in the original input. While these additions might be interpreted as hallucinations, they were consistently contextually appropriate and logically consistent with the game's setting.
Since the prompt explicitly requested additional suggestions, these cases could reflect either issues of expression or mild forms of hallucination, making their classification less clear-cut.

Another trend we observed was the model’s ability to adapt its feedback based on the completeness of the input. Game ideas that lacked specific aspects received more focused and detailed suggestions in those areas. Conversely, well-rounded ideas covering all ten aspects typically received shorter summaries, along with a few targeted improvement hints. However, these observations were only trends. Generally, the quality of the feedback varied considerably. Some responses were rich, specific, and actionable, while others were brief and more generic. This variability was sometimes influenced by the completeness and clarity of the input game idea, but not always.

%% file: files/sparc.tex
\section{SPARC}\label{sec:sparc}
\begin{figure}[h]
    \centering
    \includegraphics[trim=0 0 0 0, clip, width=0.6\textwidth]{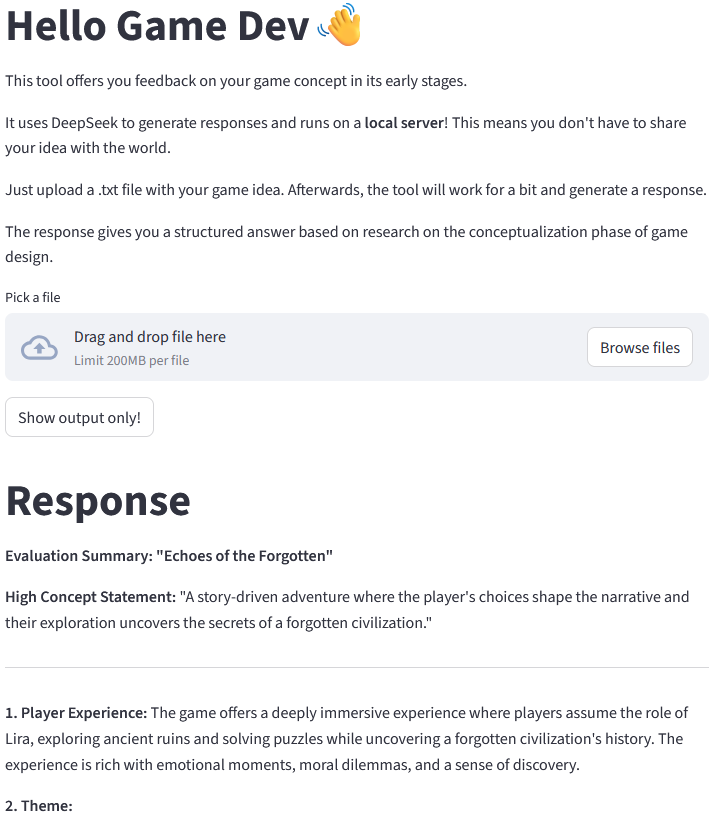}
    \caption{Screenshot of the SPARC frontend shown to participants in the study. After a text file is uploaded, the response appears at the bottom once processing is complete.}
    \label{fig:frontend}
\end{figure}

Following the selection of DeepSeek-R1, we developed a prototype tool named \textit{SPARC—System for Prototyping And Refining Concepts}—to support early-stage game design feedback in a practical setting. The tool features a minimalistic user interface built using Streamlit\footnote{\href{https://streamlit.io/}{https://streamlit.io/}} as the frontend framework (see \autoref{fig:frontend}).
In this setup, DeepSeek-R1 was integrated directly using the LangChain API\footnote{\href{https://python.langchain.com/api_reference/}{https://python.langchain.com/api\_reference/}}, with a typical response time of approximately 1–2 minutes per input on our baseline system (see \autoref{tab:system_configuration}).
SPARC allows users to upload a game concept as a plain text file to receive structured feedback directly on screen. The tool was designed to simulate real-world conditions, where users—such as students, hobbyists, or indie developers—may lack expertise in prompting or may be unfamiliar with relevant aspects like those identified in \autoref{sec:framework}.
The tool served as the central component in our pilot study, enabling us to evaluate the model’s usefulness in a context that more closely resembles actual design workflows.

\subsection{Study Design and Procedure}
With the frontend in place, we conducted a pilot user study with $n = 10$ participants. The participants were students enrolled in a narrative storytelling course jointly offered by the Technical University of Munich (TUM) and the University of Television and Film Munich (HFF). The course was structured around collaborative game development, with interdisciplinary teams of approximately four members each. Students from TUM, enrolled in the Games Engineering program, primarily focused on game programming, while HFF students came from film-related disciplines and were less directly involved in games.

\newpage
The user study took place during the early phase of the course, when teams had just developed their initial game concepts but had not yet started implementation. Participation was voluntary. In return for their time, participants received formative feedback on their game ideas generated through SPARC, which was relevant to their course project. In total, ten students across six teams took part.

The procedure was as follows:

\begin{enumerate}
    \item Participants reviewed and accepted an informed consent form.
    \item SPARC was introduced, including a brief explanation of its purpose and user interface, with emphasis on the fact that it was hosted locally.
    \item Each team submitted its initial game concept as a text file. Because SPARC was hosted on a private server at the time, we processed the files ourselves and demonstrated the results.
    \item For each team, SPARC was run once using its submitted concept. All team members were present during this process to ensure a shared understanding.
    \item The resulting outputs were distributed to participants as text files for review.
    \item Each participant then completed an individual online questionnaire, which included both closed- and open-ended questions.
\end{enumerate}

\subsection{Participants}
All ten participants were students in the joint TUM–HFF course. Their ages ranged from 22 to 31 years ($M = 25.0$, $SD = 2.6$). Reported gender was male for nine participants ($n = 9$), and one participant chose not to disclose gender. Participants reported academic backgrounds in game design, narrative, technical art, and computer science. Nine participants self-reported working in or studying game development, while the remaining participant identified primarily as a player. This contextualizes their perspective in relation to the feedback they provided.

\subsection{Results}
The results of the closed-ended questions are shown in \autoref{fig:questionaire}.
\begin{figure}
    \centering
    \begin{subfigure}{0.49\textwidth}
        \centering
        \begin{tikzpicture}[scale=0.65]
          \draw[->] (0,0) -- (6,0) node[right] {Rating};
          \draw[->] (0,0) -- (0,6) node[above] {Frequency};
          \filldraw[fill=blue!50] (0.5,0) rectangle (1.0,0);
          \filldraw[fill=blue!50] (1.5,0) rectangle (2.0,1);
          \filldraw[fill=blue!50] (2.5,0) rectangle (3.0,1);
          \filldraw[fill=blue!50] (3.5,0) rectangle (4.0,6);
          \filldraw[fill=blue!50] (4.5,0) rectangle (5.0,2);
          \node at (0.75,-0.3) {1};
          \node at (1.75,-0.3) {2};
          \node at (2.75,-0.3) {3};
          \node at (3.75,-0.3) {4};
          \node at (4.75,-0.3) {5};
          \node at (0.75,0.2) {0};
          \node at (1.75,1.2) {1};
          \node at (2.75,1.2) {1};
          \node at (3.75,6.2) {6};
          \node at (4.75,2.2) {2};
        \end{tikzpicture}
        \caption{\textit{How would you rate the quality of the response?}}
        \label{fig:quality}
    \end{subfigure}
    \hfill
    \begin{subfigure}{0.49\textwidth}
        \centering
        \begin{tikzpicture}[scale=0.65]
            \draw[->] (0,0) -- (6,0) node[right] {Rating};
          \draw[->] (0,0) -- (0,6) node[above] {Frequency};
          \filldraw[fill=blue!50] (0.5,0) rectangle (1.0,0);
          \filldraw[fill=blue!50] (1.5,0) rectangle (2.0,1);
          \filldraw[fill=blue!50] (2.5,0) rectangle (3.0,5);
          \filldraw[fill=blue!50] (3.5,0) rectangle (4.0,3);
          \filldraw[fill=blue!50] (4.5,0) rectangle (5.0,1);
          \node at (0.75,-0.3) {1};
          \node at (1.75,-0.3) {2};
          \node at (2.75,-0.3) {3};
          \node at (3.75,-0.3) {4};
          \node at (4.75,-0.3) {5};
          \node at (0.75,0.2) {0};
          \node at (1.75,1.2) {1};
          \node at (2.75,5.2) {5};
          \node at (3.75,3.2) {3};
          \node at (4.75,1.2) {1};
        \end{tikzpicture}
        \label{fig:helpfulness}
        \caption{\textit{How helpful do you think the response was?}}
    \end{subfigure}

    \vspace{1em}

    \begin{subfigure}{0.49\textwidth}
        \centering
        \begin{tikzpicture}[scale=0.65]
          \def\startA{0}
          \def\maybeA{288}
          \def\yesA{72}
          \fill[blue!30] (0,0) -- (\startA:3) arc (\startA:\maybeA:3) -- cycle;
          \fill[green!40] (0,0) -- ({\maybeA}:3) arc ({\maybeA}:360:3) -- cycle;
          \node at (90:2.2) {Maybe (80\%)};
          \node at (320:2.2) {Yes (20\%)};
        \end{tikzpicture}
        \caption{\textit{Are you likely to incorporate these suggestions into your game idea?}}
    \end{subfigure}
    \hfill
    \begin{subfigure}{0.49\textwidth}
        \centering
        \begin{tikzpicture}[scale=0.65]
          \def\startA{0}
          \def\maybeA{72}
          \def\yesA{288}
          \fill[blue!30] (0,0) -- (\startA:3) arc (\startA:\maybeA:3) -- cycle;
          \fill[green!40] (0,0) -- ({\maybeA}:3) arc ({\maybeA}:360:3) -- cycle;
          \node at (30:2.2) {Maybe (20\%)};
          \node at (210:1.5) {Yes (80\%)};
        \end{tikzpicture}
        \caption{\textit{Is this tool something that you would be interested in using again in the future?}}
    \end{subfigure}

    \vspace{1em}

    \begin{subfigure}{0.49\textwidth}
        \centering
        \begin{tikzpicture}[scale=0.65]
          \draw[->] (0,0) -- (7,0) node[right] {Rating};
          \draw[->] (0,0) -- (0,5) node[above] {Frequency};
        
          \filldraw[fill=blue!50] (0.9,0) rectangle (1.4,2);
          \filldraw[fill=blue!50] (2.5,0) rectangle (3.0,4);
          \filldraw[fill=blue!50] (4.1,0) rectangle (4.6,3);
          \filldraw[fill=blue!50] (5.7,0) rectangle (6.2,1);
        
          \node at (1.1,-0.3) {>1 min};
          \node at (2.7,-0.3) {>2 min};
          \node at (4.3,-0.3) {>5 min};
          \node at (5.9,-0.3) {>10 min};
        
          \node at (1.1,2.2) {2};
          \node at (2.7,4.2) {4};
          \node at (4.3,3.2) {3};
          \node at (5.9,1.2) {1};
        \end{tikzpicture}
        \caption{\textit{How long would be acceptable to wait for the response to be generated?}}
    \end{subfigure}

    \caption{Answer distributions for the pilot study questions. Each question is shown in the caption of the corresponding subfigure.}
    \label{fig:questionaire}
\end{figure}

In the open-ended responses, four students expressed a desire for more in-depth evaluations, suggesting that the feedback could benefit from greater detail or elaboration. One participant specifically noted that the response was \textit{"really good"} and that the tool \textit{"gave some interesting perspectives / recommendations."} However, the same participant also observed a misinterpretation in the output, where the model incorrectly identified certain elements—such as the art style—from the input.
Three students proposed an additional feature: the ability to focus on individual aspects of the game concept, rather than receiving feedback on all ten at once. Another participant suggested that the tool be made available to all students, highlighting its perceived value beyond the pilot setting.
The remaining comments were largely unrelated to the tool’s functionality—for example, non-substantive responses such as \textit{"idk"} or feedback on the naming scheme of output files.

\subsection{Discussion}
As in previous sections, this discussion focuses specifically on the outcomes of the pilot study. Broader implications are explored in \autoref{sec:discussion}.

The quantitative results presented in the accompanying figures are encouraging. Participants generally rated the quality of the model’s feedback as above average, suggesting that medium-sized LLMs like DeepSeek-R1 are capable of producing coherent and relevant responses that reflect a reasonable understanding of game concepts. However, the helpfulness of the feedback was rated slightly lower than its quality—while not negative, it indicates room for improvement in practical applicability.

Notably, two participants (20\%) indicated they would incorporate the model’s feedback into their game concepts, a modest proportion but nevertheless an increase over a baseline of zero. Moreover, 80\% of participants expressed interest in using such a tool in the future, underscoring the potential value of locally hosted systems that prioritize privacy and accessibility. This interest was further reflected in participants' willingness to tolerate longer response times, as well as in one explicit request to make the tool available more broadly to all students.

That said, limitations in functionality and output quality remain apparent. Some participants were satisfied with the feedback, while others expressed a desire for more in-depth evaluations. A commonly suggested improvement was the option to receive feedback on individual aspects rather than all ten at once. While this feature could improve focus and perceived depth, it may also increase total runtime, as it would require separate inference passes for each aspect.

There are also potential limitations in the study design that may have introduced bias. For instance, members of the same team received identical responses for their submission, influencing more than one set of answers. Additionally, participants’ game concepts might have been at slightly different stages of development and may have varied in their level of detail, which could have influenced how specific or generic the model’s responses appeared. Since we did not formally evaluate or categorize the participant-submitted ideas, we cannot control for this variable. However, based on our earlier human evaluation, it is plausible that more complete submissions (those covering most or all of the ten aspects) led to shorter or more generic feedback.

%% file: files/discussion.tex
\section{General Discussion and Implications}\label{sec:discussion}
To summarize our findings, the use of medium-sized LLMs—specifically DeepSeek-R1—on consumer-grade hardware shows considerable promise. At the upper end of the quality spectrum, the model produced strong results in both the human evaluation and the pilot study. These outcomes demonstrate the feasibility of leveraging locally hosted LLMs to support game designers without compromising privacy or intellectual property concerns.
This potential is further reflected in participants’ enthusiasm for such tools, as evidenced by their willingness to use SPARC in future projects. 

However, the results also point to clear areas for improvement. Certain aspects were better understood by the model than others, and the output continued to follow two distinct structural formats, which may affect consistency and user expectations.
A key opportunity for enhancing reliability lies in prompt engineering. Targeting individual aspects through separate prompts—an idea also suggested by participants—may improve both the specificity and depth of the feedback. Future studies could explore this structured prompting strategy in more detail.

Looking ahead, we envision an evolution of SPARC that leverages its current strength—categorizing design feedback by aspect—but moves away from generating new ideas or full rewrites. Instead, the system could help designers identify unclear or underdeveloped areas in their concepts. These areas could then be paired with targeted, thought-provoking questions drawn from a curated catalog to guide further development. This shift would support both conceptual clarity and creative iteration, aligning well with established game design practices~\cite{schell2008art, galuzin2016preproduction}.
Such an approach would transform SPARC from a general-purpose evaluator into a structured design support system, capable of helping designers not only reflect on what is missing but also how to improve it.

%% file: files/related_work.tex
\section{Related Work}\label{sec:related_work}
The optimization of language models for consumer-grade hardware is an active area of research. Xu et al.~\cite{xu2024device} provide a comprehensive review of strategies for running LLMs on resource-constrained systems across various domains. Their work highlights a range of applications—from text generation for mobile messaging~\cite{android2024gemininano} to potential use in medical diagnostics~\cite{labrak2024biomistral}. However, these applications are often geared toward general-purpose users, whereas our study distinguishes itself by focusing on supporting domain experts, specifically in game design.

The emerging field of LLM-as-a-judge explores the use of language models to assess user input or system performance~\cite{gu2024survey, li2024generation}. In a games context, Tucek et al.~\cite{tucek2024one} propose the game prototype \textit{One Spell Fits All}, in which a player’s in-game decisions are judged by LLMs for creativity and appropriateness. Their work also emphasizes running AI models locally, aligning with our interest in minimizing reliance on cloud-based solutions. However, while both approaches involve evaluating human-generated content with LLMs and prioritize local execution, our work differs in its focus: we aim to assist designers during the conceptualization phase, rather than embed AI into the gameplay loop itself.
Similarly, Hutson et al.~\cite{hutson2024enhancing} use generative AI to support assessment and feedback in game design education, with the goal of enhancing student engagement and learning outcomes. While our system also provides feedback on game concepts, our focus is not on pedagogical evaluation, but rather on helping designers iteratively refine early-stage ideas.

More broadly, the use of LLMs to support game design processes has received increasing attention~\cite{gallotta2024large, sweetser2024large}. Begemann et al.~\cite{begemann2024empirical} and Long et al.~\cite{long2024sketchar} explore the use of generative AI during the early phases of game development, emphasizing its utility in supporting creativity and concept generation. However, their focus lies primarily in visual content creation—such as image or asset generation—whereas our study focuses specifically on the textual structure and clarity of game concepts.
Lee et al.~\cite{lee2023empowering} investigate AI-supported workflows for generating complete game design proposals, including concept art and documentation, over a longitudinal study spanning four years. While we share a focus on the early stages of game development, our work differs in both scope and scale: we concentrate specifically on the written concept itself and explicitly emphasize deployment on consumer-grade hardware, making our approach more accessible to indie developers and students.

%% file: files/conclusion.tex
\section{Conclusion and Future Work}\label{sec:conclusion}
In this study, we identified ten key aspects that contribute to a strong game concept and evaluated three medium-sized language models—LLaMA 3.1, Qwen 2.5, and DeepSeek-R1—all of which can be run on consumer-grade hardware. Through a structured human evaluation, we compared the outputs of these models and selected DeepSeek-R1 for a more in-depth analysis based on its consistent formatting and coverage of the ten aspects.
Building on this, we developed SPARC (System for Prototyping and Refining Concepts), a lightweight prototype tool that enables users to upload game concepts and receive structured feedback. We then conducted a pilot study to assess SPARC’s practical effectiveness in supporting early-stage game design.
The results suggest that medium-sized LLMs are promising tools for assisting designers during the conceptualization phase, offering a balance between usability, performance, and local deployment. However, the current system still exhibits inconsistencies in output quality, and the usefulness of the feedback varies depending on the input and aspect coverage.

To address these limitations, future work will focus on improving prompt design and refining the interaction model to support aspect-specific evaluations, allowing users to request focused feedback on individual dimensions of their game concepts. Additionally, rather than offering direct suggestions—which can vary in quality—we propose an alternative strategy: generating thought-provoking questions targeting underdeveloped aspects. This approach aligns more closely with iterative design practices and aims to better support designers in refining their ideas.
By shifting from prescriptive feedback to guided reflection, future iterations of SPARC can evolve into a more effective design support system that empowers users to make meaningful improvements to their concepts.